\documentclass[12pt]{article}
\usepackage{graphicx}
\usepackage{amsmath}

\def\pbnr{}
\def\speaker{Sara Collins, Issaku Kanamori\footnote{issaku.kanamori@physik.uni-regensburg.de} and Johannes Najjar
}
\def\onbehalfof{}
\def\title{Lattice calculation\\ of\\ 
$D_s$ to $\eta^{(\prime)}$ semi-leptonic decay form factors}
\def\affiliation{%
     Institue for Theoretical Physics, University of Regensburg\\
       D-93040 Regensburg, Germany}
\def\support{}

\newcommand\pubnumber{\pbnr}
\newcommand\pubdate{\today}
\def\Title#1{\begin{center} {\Large #1 } \end{center}}
\def\Author#1{\begin{center}{ \sc #1} \end{center}}

\newcommand{\OnBehalf}[1]{\sbox0{#1}\ifdim\wd0=0pt
        {}
	\else
	{\\on behalf of #1}
	\fi}
\newcommand{\SupportedBy}[1]{\sbox0{#1}\ifdim\wd0=0pt
        {}
	\else
	{\footnote{#1}}
	\fi}
\def\Address#1{\begin{center}{ \it #1} \end{center}}

\newcommand\pubblock{\includegraphics[width=5cm]{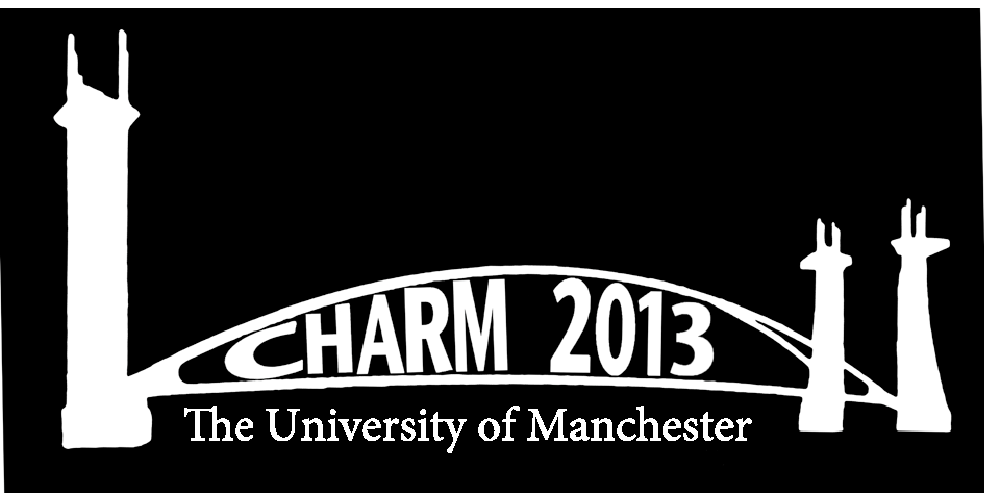}\hfill{\begin{tabular}{l} \pubnumber\\
         \pubdate  \end{tabular}}}
\newenvironment{Abstract}{\begin{quotation}  }{\end{quotation}}
\newenvironment{Presented}{\begin{quotation} \begin{center} 
             PRESENTED AT\end{center}\bigskip 
      \begin{center}\begin{large}}{\end{large}\end{center} \end{quotation}}
\def\Acknowledgements{\bigskip  \bigskip \begin{center} \begin{large}
             \bf ACKNOWLEDGEMENTS \end{large}\end{center}}
\def\venue{The 6$^{th}$ International Workshop on Charm Physics\\
(CHARM 2013)\\
Manchester, UK,  31 August -- 4 September, 2013}

\def\beq{\begin{equation}}
\def\eeq#1{\label{#1}\end{equation}}
\def\eeqn{\end{equation}}

\def\beqa{\begin{eqnarray}}
\def\eeqa#1{\label{#1}\end{eqnarray}}
\def\eeqan{\end{eqnarray}}

\let\bar=\overbar

\def\tr{{\mbox{\rm tr}}}

\def\Dslash{\not{\hbox{\kern-4pt $D$}}}
\def\dslash{\not{\hbox{\kern-2pt $\del$}}}

\def\msb{{\bar{\ssstyle M \kern -1pt S}}}

\newcommand{\DiagOnePtDisconn}[2]{%
\setlength{\unitlength}{0.1em}
\begin{picture}(36,13)(-4,6)
 \put(-4,0){\includegraphics[width=3.5em]{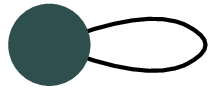}}
 \put(1.5,5.5){\small $\gamma_5$}
 \put(2,-7){#1}
 \put(22,14){#2}
\end{picture}
}

\begin{document}
\begin{titlepage}
\pubblock

\vfill
\Title{\title}
\vfill
\Author{\speaker\SupportedBy{\support}\OnBehalf{\onbehalfof}}
\Address{\affiliation}
\vfill
\begin{Abstract}
We report lattice results of $D_s$ meson semi-leptonic decay form factors 
to $\eta$ and $\eta'$ mesons.  This decay process contains disconnected 
fermion loops, which are challenging in lattice calculations.   
Our result shows that the disconnected loops give significant contributions 
to the form factors.
\end{Abstract}
\vfill
\begin{Presented}
\venue
\end{Presented}
\vfill
\end{titlepage}
\def\thefootnote{\fnsymbol{footnote}}
\setcounter{footnote}{0}
%

\section{Introduction}

Among semi-leptonic decays of charmed mesons,
decays of $D$-mesons, such as $D\to l\bar{\nu}_l K$,
are well studied both in experiment and in lattice calculations.
The high precision measurements from experiment
combined with high precision calculations of the relevant form factors 
from the lattice allows us to extract the value of the corresponding 
CKM matrix element.
When it comes to charmed mesons with a strange quark, $D_s(\bar{D}_s)$,
the situation is different, however.
The major semi-leptonic mode is $D_s \to l \bar{\nu}_l \eta^{(\prime)}$
and only the branching fractions are available from the experiment.
On the theory side, no lattice results have been published so far,
while predictions from light cone QCD sum rules 
\cite{Azizi:2010zj,Offen:2013nma} are available.
These decays are interesting for flavor physics
because of $\eta$-$\eta'$ mixing and the investigation of these
modes helps to understand the mixing angle and also possible gluonic
contributions \cite{DiDonato:2011kr}.  
For decays involving the $\eta'$,
the chiral anomaly should play some role and
it is an interesting play ground for obtaining deeper
understanding of the quantum field theory.

The relevant matrix element for these decay modes is
\begin{equation}
 \langle \eta^{(\prime)}(k)| V^\mu(q^2) |D_s(p)\rangle
 = f_+(q^2) \left[ (p+k)^\mu - \frac{M_{D_s}^2 - M_{\eta^{(\prime)}}^2}{q^2}q^\mu\right]
   + {f_0(q^2)}  \frac{M_{D_s}^2 - M_{\eta^{(\prime)}}^2}{q^2}q^\mu,
 \label{eq:vector-current}
\end{equation}
where $V^\mu$ is a vector current and $M_{D_s}$ and $M_{\eta^{(\prime)}}$ are
the masses of the $D_s$ and $\eta^{(\prime)}$, respectively.
This matrix element is characterized by two form factors, 
$f_0(q^2)$ and $f_+(q^2)$.
So far, we have focused on the scalar form factor $f_0(q^2)$,
which we can also obtain from a scalar current $S=\bar{s} c$,
\begin{equation}
 f_0(q^2) 
= \frac{m_c - m_s}{M_{D_s}^2 - M_{\eta^{(\prime)}}^2}\langle \eta^{(\prime)} |S| D_s \rangle.
 \label{eq:scalar-current}
\end{equation}
We use this relation because the combination $(m_c-m_s) S$
is protected from renormalization due to the partially conserved 
vector current \cite{Na:2010uf}.

The three point function for the above matrix element
contains
fermion disconnected loops
shown pictorially bellow (the second term is the disconnected part):
\begin{equation} 
\langle \eta^{(\prime)}(\vec{k}) | S(\vec{q}) |D_s(\vec{p}) \rangle
=
\raisebox{-1em}{\includegraphics[width=8em]{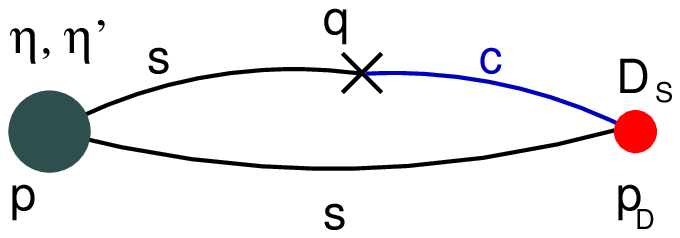}}
-\sum_{l=u,d,s}\left(
\raisebox{-1em}{\includegraphics[width=8em]{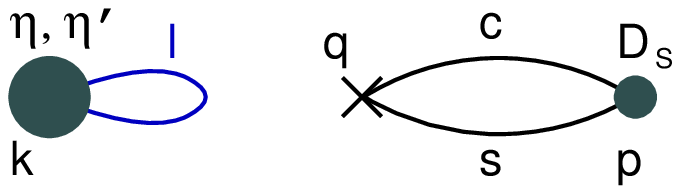}}
\right).
\label{eq:3pt}
\end{equation}
Here, only the valance quark lines are drawn
and all possible gluon lines and sea quark loops are suppressed.
The calculation of disconnected loops is expensive and the signal is
noisy (but feasible \cite{Bali:2011yx}),
so if the contributions were small it might be 
a good approximation to neglect them.  
This is not the case for the matrix element above, however.
The disconnected loops should be summed over three light flavors
which enhances the magnitude roughly by a factor three.
They contain the contributions from the chiral anomaly to the $\eta'$-meson.
For these reasons,
the disconnected loops may contribute significantly.

\section{Lattice Setup}
We use QCDSF $n_f=2+1$ configurations~%
\cite{Bietenholz:2010jr, Bietenholz:2011qq},
which were generated with
the tree level Symanzik improved gluon action
and non-perturbative stout link improved clover fermion action.
Although we neglect charm sea quarks but included relativistic
valance charm quark.
The charm quark mass was tuned to reproduce the spin averaged
physical charmonium mass $M_{\rm \overline{1S}}$.
The $u$-, $d$- and $s$- quark masses are tuned so that their average
$\frac{1}{3}(m_u+m_d+m_s)$ is fixed 
and that $2M_K^2 +M_\pi^2$ coincides with the physical values.
Due to this strategy of choosing light quark masses,
these configurations are particularly
suitable for studying flavor physics using the SU(3) flavor basis.
So far we have analyzed two sets of configurations,
one is at the SU(3) flavor symmetric point ($m_{u,d}=m_s$) 
with $M_\pi \simeq 450\, {\rm MeV}$
($N_{\rm conf}=939$), and the other has lighter $m_{u,d}$
and heavier $m_s$, which gives $M_\pi \simeq 348\, {\rm MeV}$ 
($N_{\rm conf}=239$).
Both sets have the lattice spacing $a\sim 0.08\, {\rm fm}$,
and the lattice size $24^3\times 48$ gives the physical extent 
$L \sim 1.9\,{\rm fm}$.

The most computationally 
expensive quantity to calculate is the fermion one point loop,
\begin{equation}
  \mathcal{C}_{\rm 1pt}^f(t,\vec{k})
 = \sum_{\vec{x}}\exp(i\vec{k}\cdot\vec{x})\, \tr\! \left[
     \sum_{\vec{x}',\, \vec{x}''}
    \gamma_5 \phi(\vec{x},\vec{x}'') M^{-1}_f(t,\vec{x}''; t,\vec{x}')
    \phi(\vec{x}',\vec{x})
     \right]
 = \DiagOnePtDisconn{$t,\vec{k}$}{$f$},
 \label{eq:1pt-def}
\end{equation}
where $M_f$ is the Dirac operator for flavor $f$ and 
$\phi(\vec{x}, \vec{x}')$ is a smearing function to make 
the wave function have finite spacial extent.
Essentially, this is the trace of the inverse Dirac operator, which is 
a $O(10^7)\times O(10^7)$ matrix.
A direct calculation is practically impossible.
We use a stochastic method, which relies on an approximate completeness
relation of random noise vectors 
$|n_i\rangle$:
$\frac{1}{N}\sum_{i=1}^N |n_i\rangle \langle n_i| = 1 + O(\frac{1}{\sqrt{N}})$.
To reduce the error from the $O(\frac{1}{\sqrt{N}})$ term,
we combined several noise reduction techniques \cite{Bali:2009hu,Bali:2011yx}.

\section{Extracting $\eta$ and $\eta'$ states}

To calculate the matrix element (\ref{eq:3pt}),
we first need to build the interpolating operators 
to create the $D_s$ meson and annihilate the $\eta^{(\prime)}$ meson.
The operator for $D_s$ is easy to obtain but
ones for $\eta^{(\prime)}$ are non-trivial due to the mixing.

We start with SU(3) octet-singlet basis, 
$\eta_8=\frac{1}{\sqrt{6}}\left( u\bar{u} + d\bar{d} -2 s\bar{s}\right)$
and
$\eta_1=\frac{1}{\sqrt{3}}\left( u\bar{u} + d\bar{d} + s\bar{s}\right)$.
We can build the following $2 \times 2$ two-point correlation function
by using smeared interpolating operators, 
$\mathcal{O}_8$ for the octet and $\mathcal{O}_1$ for the singlet:
\begin{equation}
 C_2(t;\vec{k})=
 \begin{pmatrix}
  \langle \mathcal{O}_8(t;\vec{k})\mathcal{O}_8^\dagger(0) \rangle
  & \langle \mathcal{O}_8(t;\vec{k})\mathcal{O}_1^\dagger(0) \rangle
  \\
  \langle \mathcal{O}_1(t;\vec{k})\mathcal{O}_8^\dagger(0) \rangle
  & \langle \mathcal{O}_1(t;\vec{k})\mathcal{O}_1^\dagger(0) \rangle
 \end{pmatrix}.
 \label{eq:2pt-2x2}
\end{equation}
Here, $t$ is the sink-source time separation and the operator at the
sink is projected to momentum $\vec{k}$.
Each element of eq.~(\ref{eq:2pt-2x2}) contains both connected 
and disconnected fermion contributions.  

\begin{figure}
 \includegraphics[width=0.48\linewidth]{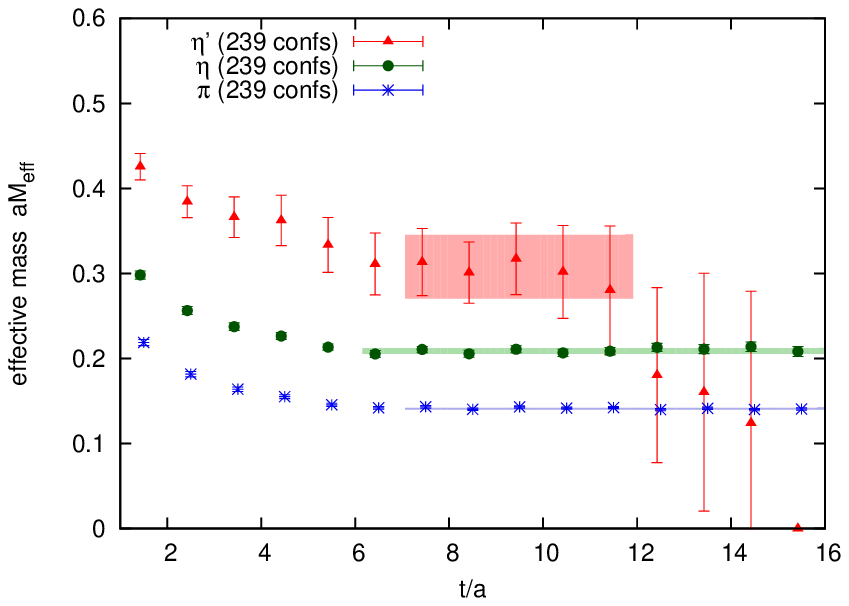}
 \hfil
 \includegraphics[width=0.48\linewidth]{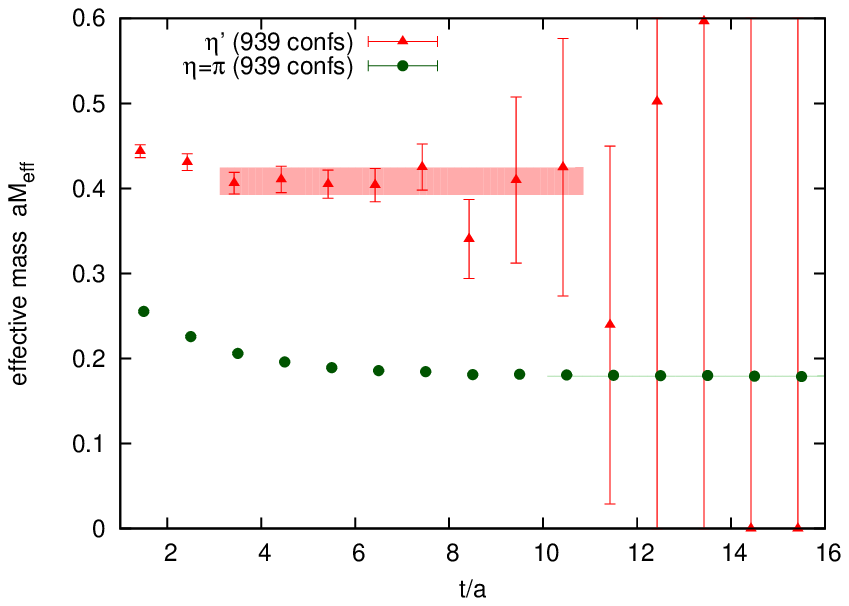}
 \vspace*{-0.5em}

\caption{Effective mass plot of $\eta'$ and $\eta$. 
For comparison, the effective mass for $\pi$ is also plotted.  
Left panel: $M_\pi = 348\, {\rm MeV}$ case.
Right panel: $M_\pi =450\, {\rm MeV}$ case (SU(3) flavor symmetric),
where $M_\pi=M_\eta=M_{\eta_8}$ and $M_{\eta'}=M_{\eta_1}$.
}
\label{fig:effmass}
\end{figure}

Two point functions in the physical $\eta$ and $\eta'$ basis
are obtained by diagonalizing eq.~(\ref{eq:2pt-2x2}).
The eigenvectors, which can be parameterized one mixing angle $\theta$,
give the interpolating operators for $\eta$ and $\eta'$:
\begin{align}
 \mathcal{O}_{\eta}
 & = \cos\theta\, \mathcal{O}_8 - \sin\theta\,  \mathcal{O}_1,
 &
 \mathcal{O}_{\eta'}
 & = \sin\theta\, \mathcal{O}_8 + \cos\theta\,  \mathcal{O}_1.
 \label{eq:O_eta-O_eta'}
\end{align}
The mixing angle above corresponds to a mixing between 
the interpolating operators
and is not the mixing of the physical observables.
The resulting two point correlators 
have the functional form:
\begin{equation}
 C_{\rm 2pt}^{\eta^{(\prime)}}(t,\vec{k})
 =\frac{|Z_{\eta^{(\prime)}}(\vec{k})|^2}{2E_{\eta^{(\prime)}}(\vec{k})}
  e^{-E_{\eta^{(\prime)}}(\vec{k})t} 
+ (\text{contribution from excited states}).
\end{equation}
We can obtain the energy $E_{\eta^{(\prime)}}$ and overlapping factor
$Z_{\eta^{(\prime)}} 
= \langle \eta^{(\prime)} | \mathcal{O}_{\eta^{(\prime)}} |0\rangle$
by fitting the two point function.  We can even obtain the energy gap
to the first excited state in some cases.
Fig.~\ref{fig:effmass} shows the effective mass 
$aM_{\rm eff}(t+\frac{a}{2}) = -\ln\frac{C_{\rm 2pt}(t+a)}{C_{\rm 2pt}(t)}$ 
(or $\frac{C_{\rm 2pt}(t+a)}{C_{\rm 2pt} (t)}
 = \frac{\cosh[ aM_{\rm eff}(t+a/2)(t+a-T/2)]}{\cosh[aM_{\rm eff}(t+a/2)(t-T/2)]}$
with the finite temporal size $T$)
at $\vec{k}=\vec{0}$, together with the fitted value of the mass.

\section{Results}

\begin{figure}
 \includegraphics[width=0.49\linewidth]{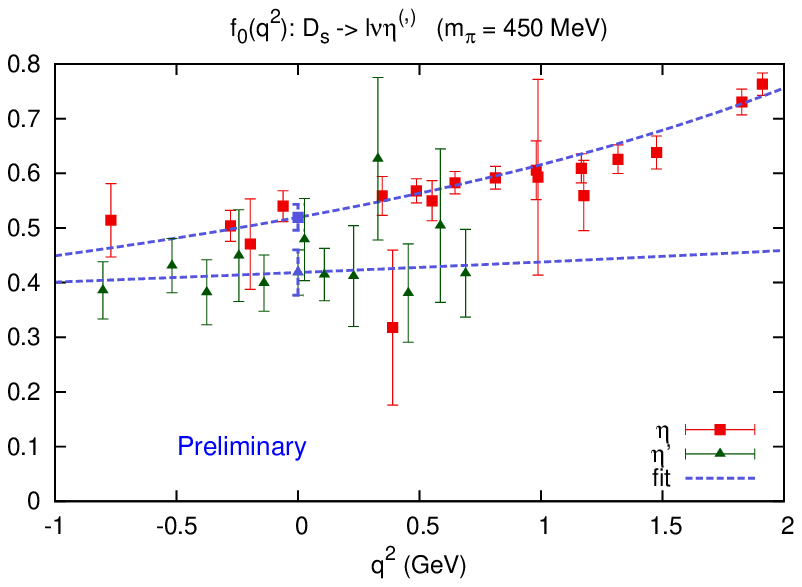}
 \hfil 
 \includegraphics[width=0.49\linewidth]{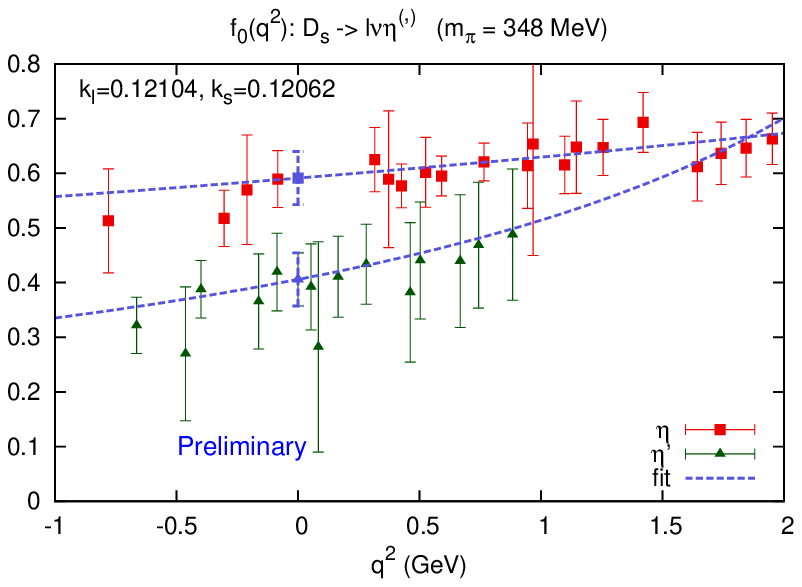}
 \vspace*{-0.5em}

 \caption{Preliminary results for the scalar form factor $f_0(q^2)$,
at the SU(3) flavor symmetric point (left panel) and 
$M_\pi=348\, {\rm MeV}$ point (right panel).
}
 \label{fig:f0}
\end{figure}

Having obtained the interpolating operators~(\ref{eq:O_eta-O_eta'}),
we can construct the three point function needed for the form factor:
\begin{align}
 C_{\rm 3pt}(t,\vec{p},\vec{q},\vec{k}) 
 &= \langle 0| \mathcal{O}_{\eta(')}(\vec{k}, t_{\rm sep}) S(\vec{q},t)
 \mathcal{O}_{D_s}^\dagger(\vec{p},0) | 0 \rangle \nonumber\\
 &= \frac{Z_{\eta(')}}{2E_{\eta(')}} \frac{Z_{D_s}}{2E_{D_s}} 
    \exp\left[-E_{D_s}t - E_{\eta(')}(t_{\rm sep} -t)\right]
 \times \Bigl[  
  {\langle \eta^{(\prime)}(\vec{k}) |S(\vec{q}) |D_s(\vec{p})\rangle} 
   + \cdots \Bigr],
\end{align}
where $t_{\rm sep}$ is the sink-source separation.
The factors in front of the matrix element 
$\langle \eta^{(\prime)}(\vec{k}) |S(\vec{q}) |D_s(\vec{p})\rangle$
can be obtained by fitting the two point functions.
Thus, we can obtain the scalar form factor $f_0(q^2)$
by using eq.~(\ref{eq:scalar-current}).
The results are plotted in Fig.~\ref{fig:f0},
where the errors were calculated with jackknife analysis.
We fit the results with one pole functions, $f_0(q^2)= f_0(0)/(1-bq^2)$.
The preliminary values at $q^2=0$ are
$f_0^{D_s\to\eta}(0)=0.52(2)$ and 
$f_0^{D_s\to\eta'}(0)=0.42(4)$ for the SU(3) symmetric case,
$f_0^{D_s\to\eta}(0)=0.59(5)$ and 
$f_0^{D_s\to\eta'}(0)=0.41(5)$ for the $M_\pi=348\,{\rm MeV}$ case.

\begin{figure}
 \begin{minipage}[t]{0.47\linewidth}
  \hspace*{-0.05\linewidth}
  \includegraphics[width=1.05\linewidth]{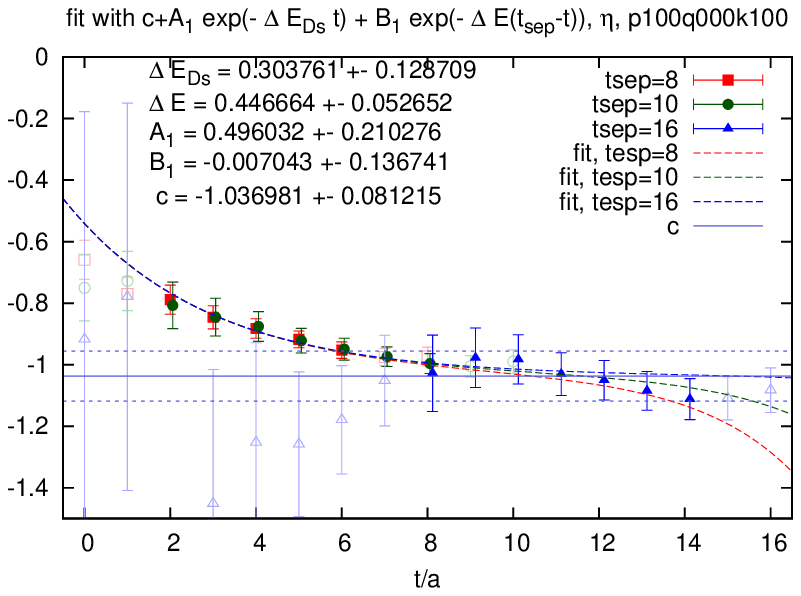}
 \vspace*{-1.5em}

  \caption{Fit of $R(t)$ with excited contributions from $D_s$ and
  $\eta$.  $D_s\bigl(\vec{p}=(1,0,0))\bigr)$ is located at $t/a=0$
  and $\eta\bigl(\vec{k}=(1,0,0)\bigr)$ is at $t/a=8,\ 10$ and $16$.
  The data points with open symbols were not used in the fitting.
  }
  \label{fig:fitting_ratio}
 \end{minipage}
\hfil
 \begin{minipage}[t]{0.47\linewidth}
  \hspace*{-0.05\linewidth}
  \includegraphics[width=1.05\linewidth]{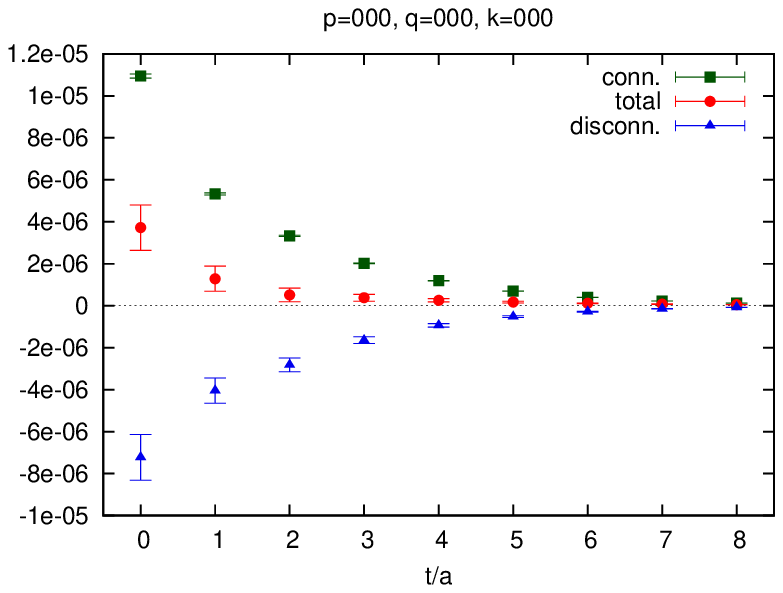}
 \vspace*{-1.5em}

  \caption{The connected, disconnected, and total contributions to 
  $C_{\rm 3pt}(t)$ with $t_{\rm sep}=8$ 
  for $D_s \to l\bar{\nu}\eta'$ at SU(3) symmetric point. 
  $D_s$ and $\eta'$ are locate at $t/a=0$ and $8$, respectively.}
  \label{fig:conn_disconn}
 \end{minipage}
\end{figure}

To extract the matrix element between the ground state of $D_s$ and 
$\eta^{(\prime)}$, it is important to remove the excited contributions.
After removing the leading exponential factors by taking a ratio,
we have
\begin{align}
R(t)
 &\equiv
 \frac{C_{\rm 3pt}(t,\vec{p},\vec{q},\vec{k})}
{\frac{Z_{\eta(')}}{2E_{\eta(')}} \frac{Z_{D_s}}{2E_{D_s}} 
    \exp(-E_{D_s}t - E_{\eta(')}(t_{\rm sep} -t))}
 \nonumber\\
&= \langle \eta^{(\prime)}(\vec{k}) |S(\vec{q}) |D_s(\vec{p})\rangle
    + A_1 \exp(-\Delta E_{D_s} t) + B_1 \exp(-\Delta E_{\eta^{(\prime)}}(t_{\rm sep} -t))
   + \cdots, 
\label{eq:ratio}
\end{align}
where $\Delta E_{D_s}$ and $\Delta E_{\eta^{(\prime)}}$ are the energy
gaps to the first excited state.
If $t_{\rm sep}$ and $t$ were large enough,
the residual pollution from the excited states would be exponentially small 
and negligible.  
We can not use such large $t_{\rm sep}$ and $t$, 
however, because the statistical error would increase
and we would not obtain meaningful signals.
We instead fit the ratios (\ref{eq:ratio}) 
with three terms in the r.h.s.\ simultaneously with three different 
$t_{\rm sep}/a=(8,10,16)$ (for $\eta$ in the SU(3) symmetric point, 
we also used $t_{\rm sep}/a=24$), 
using the energy gaps obtained from the two point functions.
If we were not able to extract $\Delta E_{\eta^{(\prime)}}$ from the two
point function,
we only used the first two terms in eq.~(\ref{eq:ratio}). 
A typical example of the fitting is shown in
Fig.\ref{fig:fitting_ratio}.
The plot shows that the ratio is well described with the fit function.

It is interesting to note that the disconnected contributions
are really large in the $D_s \to \eta'$ three point function
(Fig.~\ref{fig:conn_disconn}).
The disconnected part has almost the same magnitude as the connected
part but the opposite sign, so the total three point function
becomes much smaller in the magnitude than the connected contribution.
Fig.~\ref{fig:conn_disconn} also implies that the error is mainly from
the disconnected part.

\section{Conclusions}

We gave the preliminary results of a lattice calculation of
semi-leptonic decay form factors
for $D_s \to l\bar{\nu} \eta$ and $D_s \to l \bar{\nu} \eta'$,
including the disconnected fermion loop contributions.
This is the first lattice result for these form factors.
The disconnected contributions to the decay $D\to l\bar{\nu}\eta'$
are significant.
the decay into $\eta'$.
We were able to obtain the scalar form factor $f_0(q^2)$ 
at $q^2=0$ in 10--15\% statistical error.

We are planning to calculate the other form factor $f_+(q^2)$,
which requires a renormalization factor.  Another aim is to
calculate form factors decaying into $\phi$, for which a rigorous 
treatment requires disconnected contributions.
The mixing of $\eta$ and $\eta'$, and its quark mass dependence
are also interesting.

\Acknowledgements

This work was supported by the DFG (SFB/TRR 55) and the EU (ITN STRONGnet).
We used a modified version of the CHROMA software suite \cite{Edwards:2004sx}.
We used time granted by PRACE at Fermi in CINECA,
as well as the Athene HPC cluster and iDataCool
at the University of Regensburg. 
We thank
G. S. Bali, S. D\"urr, B. Gl\"a\ss{}le, E. Gregory, C. McNeile,
P. P\'erez-Rubio and A. Sch\"afer for discussions.

\end{document}